\documentclass[twocolumn,preprintnumbers,amsmath,amssymb]{revtex4}
\usepackage{graphicx}
\usepackage{dcolumn}
\usepackage{bm}

\begin{document}
\title{A proposed new route to d$^0$ magnetism in semiconductors}
\author{J. Berashevich}
\address{Thunder Bay Regional Research Institute, 290 Munro St., Thunder Bay, ON, P7B 5E1, Canada}
\address{Max Planck Institute for the Physics of Complex Systems, Nöthnitzer Str. 38, 01187 Dresden, Germany}
\author{A. Reznik}
\address{Thunder Bay Regional Research Institute, 290 Munro St., Thunder Bay, ON, P7B 5E1, Canada}
\address{Department of Physics, Lakehead University, 955 Oliver Road, Thunder Bay, ON, P7B 5E1}

\begin{abstract}
Here we propose to induce magnetism in semiconductor utilizing the 
unique properties of the interstitial defect to act as the magnetic impurity 
within the $\alpha$-PbO crystal structure. The Pb$_i$ interstitial 
generates the $p$-localized state with two on-site electrons
to obey the Hund's rule for their ordering. It is demonstrated that instead of Pb interstitial 
other non-magnetic impurities of $s^2p^{x}$ outer shell configuration can be applied to 
induce d$^0$ magnetism with possibility to tune the local magnetic moments $\mu_B$ 
by varying a number of electrons $1\leq x\leq 3$.
The magnetic coupling between such defects is found to be driven by the long-range
order interactions that in combination with high defect solubility promises the 
magnetic percolation to remain above the room temperature.
\end{abstract}

\maketitle
Spintronics has recently emerged as a widely successful technology that
exploits the principles of magnetism, but the classical metal ferromagnets are not applicable there.
That has led to the creation of a new branch of research directed at semiconductors
exhibiting magnetism at room temperature \cite{ando}. The first success in this regard was with
the so-called "diluted magnetic semiconductors" created by doping semiconductors
with magnetic ions whose inner $3d$ or $4f$ shells being partially filled
that allows the ferromagnetic spin alignment \cite{prel,coey,mah,zunger}.
A keen interest on ferromagnetic semiconductors also arose when 'non-magnetic' materials
were discovered to demonstrate magnetism \cite{khan,venk,podila,dev}. Origin of so-called $d^0$ magnetism
was proposed to be due to localized $sp$ states: vacancies create a network of
unpaired electrons and as interaction between such defects can demonstrate the long-range order,
their magnetic coupling occurs \cite{dev}.

Here we propose to induce $d^0$ magnetism by forming the interstitial defect Pb$_i$ 
(Pb$_i$:$6s^26p^2$ outer shell) within the crystal structure of the tetragonal lead oxide $\alpha$-PbO.
It generates the state occupied by two $p$-unpaired electrons Pb$_i$:$6p^2$. 
Ordering of the localized electrons Pb$_i$:$6p^2$ obeys the first Hund's rule thus resulting in
formation of the stable local magnetic moment of 2.0 $\mu_B$.
The spin-polarization energy of such state is $E_{M}= E_{AFM}-E_{FM}$=0.235 eV,
where $E_{AFM}$ and $E_{FM}$ are the total energies of the anti-(AFM) and ferromagnetic states (FM), respectively \cite{exp_my}.
Therefore, otherwise than magnetism occurs due to the partially filled $p$ shell instead of the $3d$ or $4f$ shells,
Pb$_i$ works exactly as the magnetic impurity. It demonstrates
the high on-site spin stability, its $p$ electrons are localized on impurity site showing 
a weak perturbation with the host (an interaction with the host occurs through Pb$_i$:$6s^2$ electrons)
and in addition, the interstitial defect is almost non-invasive 
to the electronic and crystal structures of the host.
We have established that any impurity of the $s^2p^{1\leq x\leq 3}$
outer shell embedded as the interstitial defect into the $\alpha$-PbO crystal structure
would act as magnetic. The $p$-shell occupation, $p^{1\leq x\leq 3}$,
can be used to control the spin ordering between impurities 
and also to tune the local magnetic moments: $x=2$ works for 2.0 $\mu_B$, while $x=1$ or $x=3$ for 1.0 $\mu_B$.
From the technological point of view, the layered crystal structure of $\alpha$-PbO
promises the superior advantages in achievement of the magnetic percolation and more importantly,
a practical way for its control: {\it i)} the crystal structure of PbO type grows
in polycrystalline form whose large surface area offers enormous potential for doping 
(on surface the impurities can be placed on the nearest-neighbours); 
{\it ii)} the dopant solubility and long-range order interactions can be manipulated with 
the impurity atomic radius; {\it iii)} any compounds of the tetragonal
PbO type can be used as a semiconductor matrix.

In our study we applied the generalized gradient approximation
(GGA) with the PBE parametrization \cite{perdew} provided by WIEN2k
package for the density functional calculations \cite{blaha}.
The supercell approach ($RK_{max}$=7) with sufficiently large supercell
of 108-atom size (3$\times$3$\times$3 array of the primitive unit cells) for single impurity and of
160-atom size (5$\times$4$\times$2) for two interacting impurities have been used.
The Pb:$5p,5d,6s,6p$ and O:$2s,2p$ electrons have been treated as the valence electrons.
For integration of the Brillouin-zone, the Monkhorst-Pack scheme using a (5$\times$5$\times$4) k-mesh was applied.
The localization of the impurity
wavefunction has been additionally examined with HF applied directly to the unpaired electrons 
that allows to preserve accuracy provided by DFT but in the same time
to correct the unpaired electrons self-interaction \cite{Avezac}.

The tetragonal lead oxide $\alpha$-PbO possesses
the layered structure leading to formation of platelets 
upon compound growth and each platelets is considered as a single crystal.
The layers within such crystal are held together by the interlayer interaction of Pb:$6s^2$ electrons \cite{interlayer}. 
In terms of electronic properties, these interactions 
induce a deeping of the conduction band at $\mathbf{M}^*$ point \cite{masses} and the gap shrinks as interactions enhance. 
GGA tends to overestimate the interlayer separation in the layered structures but 
in the same time it also well known to underestimate the band gap size. 
For $\alpha$-PbO system it results in compensation effect \cite{defects}:  
for the lattice parameters optimized 
with GGA the band gap is only slightly 
underestimated as 1.8 eV against the experimental value of 1.9 eV \cite{exp}
(application of the experimental value of the lattice parameters causes the band gap to shrink 
by 0.22 eV). The correct band gap size is important in investigation of the magnetic impurities:
when the band gap size is underestimated
the hole-carrying impurity orbital occurs closer to the 
the host conduction band in turn forcing a delocalization of the defect tails.
In order to avoid the spurious long-range order interactions
\cite{zunger} all band structure calculations 
have been performed for the lattice parameters optimized with GGA. However, 
since an incorporation of the interstitial becomes easier 
as interlayer distance increases,
the interlayer distance has also crucial impact on the defect formation energy. 
In order to make correct evaluation of the formation 
energy, the experimental value of the lattice parameters has been applied for those calculations.

\begin{figure}
\includegraphics[scale=0.19]{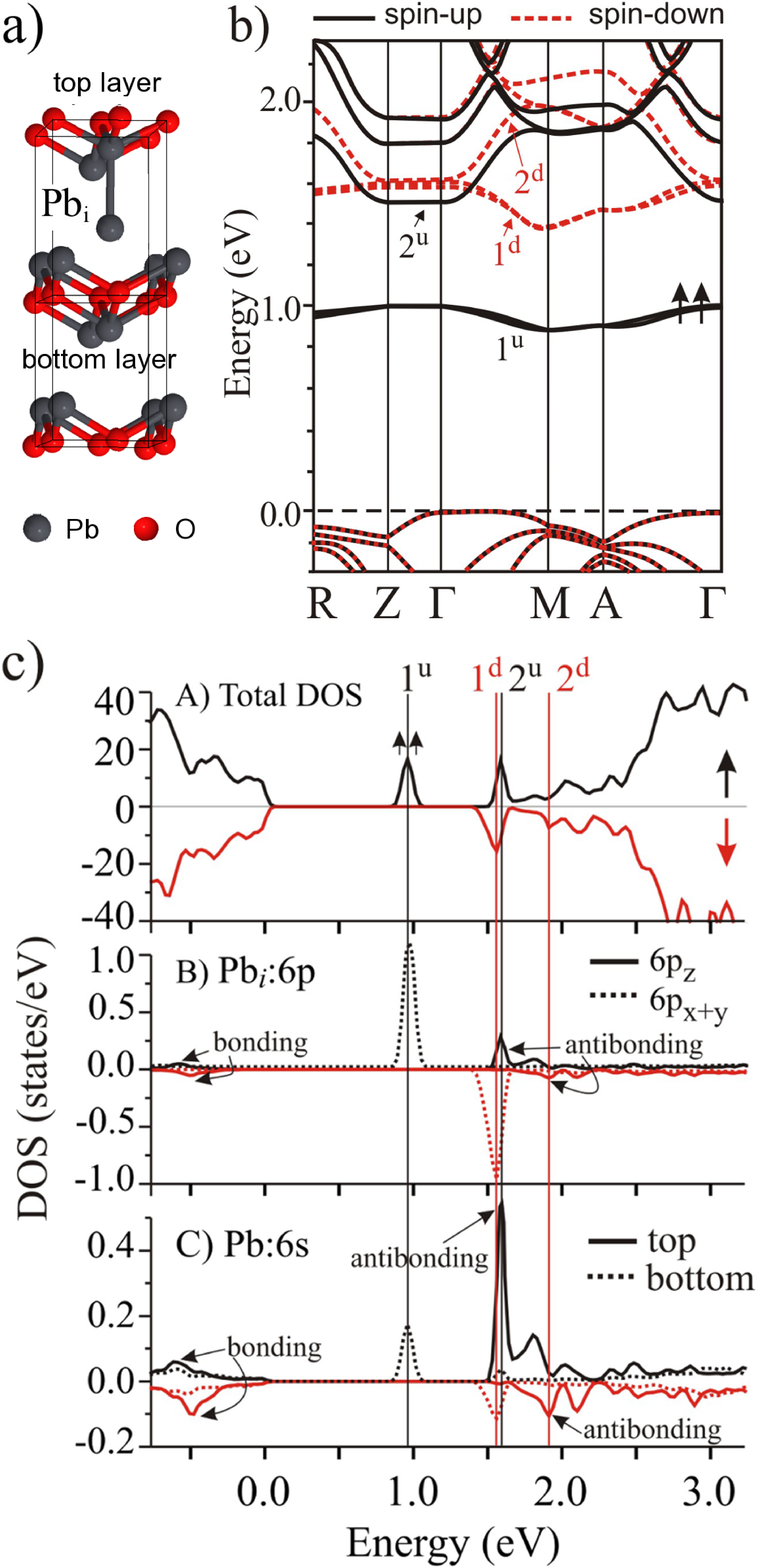}
\caption{\label{fig:fig1} (a) The Pb interstitial 
in the layered crystal structure of $\alpha$-PbO
(b) Band diagram: $1^u$ and $1^d$ are for the defect induced bands,
while $2^u$ and $2^d$ are the antibonding orbitals of the Pb$_i$-Pb bond.
(c) Total and partial density of states for Pb$_i$ and host Pb atoms
('top' is referred to Pb atom forming the Pb$_i$-Pb bond, while 'bottom'
for Pb atoms in the bottom layer). The 'bonding' and 'antibonding' in panels B and C are
reffered to the Pb$_i$-Pb bond.}
\end{figure}

The interstitial defects have never been considered in origin of 
d$^0$ magnetism because they do not belong to a class of most
common defects in crystalline materials as they induce a
significant perturbation of the host lattice resulting in their large formation energy \cite{janotti}.
The layered structure of $\alpha$-PbO is different, as it allows the foreign atoms to squeeze
between layers inducing a minimal lattice deformation to the host immediate
neighborhood (see Fig.~\ref{fig:fig1} (a)).
As a result, the formation energy of Pb$_i$ is not too high.
For the Pb rich conditions/vacuum the neutral charge state is characterized by the formation energy
of 1.23 eV (for methods see \cite{defects}).

The band diagram and the density of states (DOS) for $\alpha$-PbO containing Pb$_i$ are
presented in Fig.~\ref{fig:fig1} (b) and (c), respectively. The Pb interstitial makes a
bond with Pb atom from the host generating the defect state inside the band gap.
The defect-induced spin-up and spin-down bands (1$^u$ and 1$^d$ in Fig.~\ref{fig:fig1} (b) and (c))
both show dominant Pb$_i$:$6p_{x+y}$ character but demonstrate a different behavior.
The spin-up band is filled with two electrons and appears just above the midgap at 0.99 eV + $E_V$,
where $E_V$ is a top of the valence band (VB). In contrast, the spin-down band
(1$^d$) is unoccupied and appears at the edge of the conduction band (CB).
In order to understand an asymmetry in filling of the spin-up and spin-down states,
bonding of Pb$_i$ with the host lattice has to be considered.

The integration of the Pb interstitial into the crystal lattice
requires excitation of its ground state $6s^26p^2$ to $6s^16p^3$. 
The Pb$_i$:$6s^1$ and Pb$_i$:$6p^1_{z}$ electrons participate in bond formation with the
Pb atom from the host which shares its Pb:$6s^2$ electrons (see Fig.~\ref{fig:fig1} (c)).
The bonding orbital appears inside VB, while the antibonding forms the CB bottom
(see $2^u$ and $2^d$ bands and notations of 'bonding' and 'antibonding' at B and C panels to DOS).
The length of the Pb$_i$-Pb bond is fairly short, 2.9 \AA, that is indicative of a double bond formation.
An out-of-plane displacement
of 0.54 \AA\ occurs for the Pb atom involved in bonding.
Because participation of the host Pb:$6s^2$ orbitals in bonding interferes with 
interlayer interactions, it alters a behavior of the conduction 
bands near the CB top: in analogy with the band behavior in a single layer \cite{masses},
the band deeping at the $\mathbf{M}^*$ point is suppressed.

Therefore, out of three Pb$_i$:$6p^3$ electrons found in the excited state ($6s^16p^3$),
only one Pb$_i$:$6p^1_{z}$ participates in bonding while the two others Pb$_i$:$6p_{x+y}$
are left at the defect site.
The Hund's rule dictates both unpaired electrons to occupy the $1^u$ spin-up band.
The empty $1^d$ spin-down band is pushed up to CB that causes considerably
large spin-exchange splitting of order 0.523 eV.
The ferromagnetically ordered Pb$_i$:$6p_{x+y}$ electrons generate the local magnetic moment of 2.0 $\mu_B$.
The difference in total energy between the AFM and FM states is found to be
$E_{M}=E_{AFM}-E_{FM}$=0.235 eV thus promising the FM state to be stable well above the room temperature.
An application of the HF approach to the unpaired electrons induces 
further stabilization of the FM state to $E_{M}$=0.490 eV due to stronger on-site localization of the impurity 
wave function (the $1^u$ orbital is shifted towards the valence band 
by 0.5 eV thus enhancing the splitting of the $1^u$ and $1^d$ orbitals).
On the other hand, reduction of the interlayer distance (the experimental lattice parameters are applied)
in opposite causes delocalization of the wave function due to stronger hybridization of impurity state with the 
host lattice.

The $6s^26p^2$ electronic configuration is a key point for the
Pb interstitial to act as a magnetic impurity: filled $s^2$ outer shell is required
for bonding with the host lattice while partially filled $p$ shell contributes
in development of the local magnetic moment. Following this principle, other chemical
elements possessing $s^2p^{1\leq x\leq 3}$ outer shell can be applied to induce $d^0$
magnetism. Indeed, several examined impurities have shown a formation 
of the local magnetic moment,
impurity of $x=1$ or $x=3$ induces the local magnetic moments of 1 $\mu_B$, while $x=2$ produces 2 $\mu_B$.
The stability of the spin-polarized state is observed to grow substantially
with reduction of the impurity atomic radius. 

\begin{figure}
\includegraphics[scale=0.40]{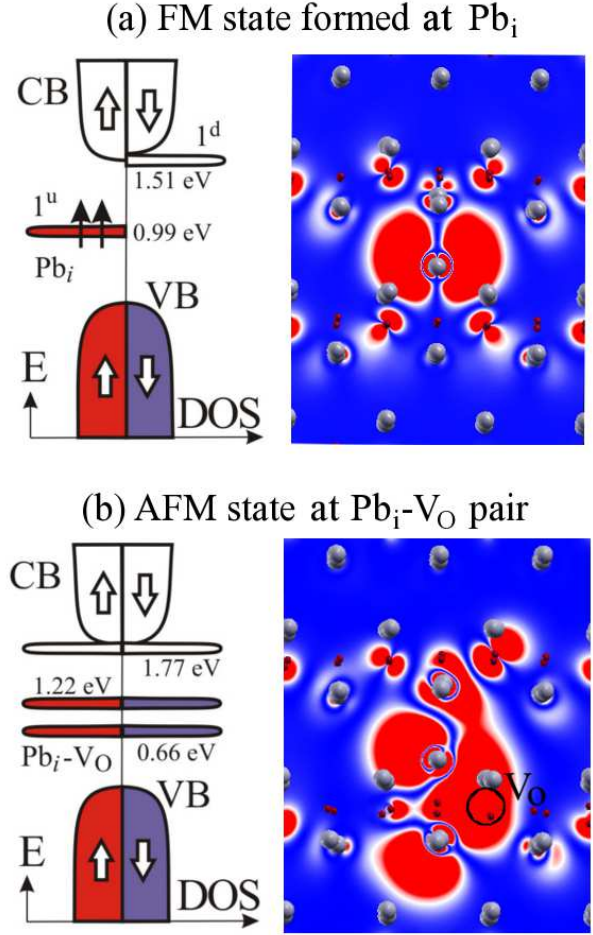}
\caption{\label{fig:fig2} Schematic band diagrams demonstrate a formation of
FM (Pb$_i$) and AFM (Pb$_i$-$V_{\operatorname{O}}$ pair) states. The energy values are 
appearance of the defect levels relative to $E_V$. The electron density map is plotted
for isovalues of $\pm$0.005 e/\AA$^3$ with help of Xcrysden.
(a) FM state at the Pb$_i$ site.
The density map demonstrates the spin density calculated for the energy range
(1.0 eV+$E_V$)$\pm$0.15 eV.
(b) An interaction of Pb$_i$ with $V_{\operatorname{O}}$ destroys the FM state at the Pb$_i$ site.
The electron density is plotted for the energy range (0.9 eV+$E_V$)$\pm$0.4 eV.}
\end{figure}

Our studies of localization of the electron density at the defect site
has revealed a duality in its behavior.
Since occupied band $1^u$ is located deep inside the band gap, the
Pb$_i$:$6p^2_{x+y}$ electrons shows strong localization on the defect site thus
allowing to form the stable local magnetic moment of 2$\mu_B$ (see Fig.~\ref{fig:fig2} (a)).
However, the long defect tails appear as well but due to bonding/hybridization
of the Pb$_i$:$6s^1$ and Pb$_i$:$6p^1_{z}$ electrons from the Pb interstitial 
and Pb:$6s$ and Pb:$6p$, O:$2p$ electrons from the host (see partial DOS in Fig.~\ref{fig:fig1} (c)).
As a result, the electron density around the defect site is spin-polarized 
showing an anisotropy in its distribution: the defect tails are
symmetrically polarized relative to Pb$_i$ thus contributing in stabilization of the ferromagnetic state.

Since extension of the wave function tails to the host lattice seems 
to be of the long-range order, it is expected to induce effective defect-defect interactions. In this respect we have 
investigated an interaction of Pb$_i$ with the O vacancy ($V_{\operatorname{O}}$) which formation is also 
favored at the Pb-rich/O-poor limit (oxygen deficiency).
In its charge neutral state, $V_{\operatorname{O}}$ is occupied by two electrons
showing the strong localization at the vacancy site \cite{defects}.
Since the virtual hopping is allowed between the states formed by Pb$_i$ and $V_{\operatorname{O}}$,
they can work as compensating centers to each other.
For the defects appearing at the nearest-neighbour sites,
the FM ordering of the localized spins at Pb$_i$ is found
to be destroyed due to the strong interaction between defects to be accompanied by hybridization.
The Pb$_i$:$6p_{x+y}$ electrons from the interstitial and the Pb:$6s$ and
Pb:$6p$ and O:$2p$ electrons from the host lattice, all have been found
to contribute equally in formation of both defect states
(see manifestation of the electron density in the inter-defect region in Fig.~\ref{fig:fig2}(b)).
The strong interactions between defects contribute in lowering
of the formation energy of defect pair by 0.84 eV in comparison to 
the non-interacting defects. With defect separation the electronic interactions
are suppressed leading to recovery of the FM state: 
for distance of 6.2 \AA\ between the defects the local magnetic moment is 1.8 $\mu_B$.
However, because interaction of Pb$_i$ with $V_{\operatorname{O}}$ do not vanish completely,
the magnitude of the spin polarization energy is found to be considerably low $E_{M}=E_{AFM}-E_{FM}$=0.06 eV.
Therefore, the electronic interactions between Pb$_i$ and $V_{\operatorname{O}}$
demonstrate the long-range order due to the long tails of the Pb$_i$ interstitial. 

For two interacting interstitial defects, as both defects contribute
with their tails into the long-range order interactions, the magnetic coupling between them
is expected to sustain over the larger distance. 
The $6p^2_{x+y}$ state is exactly half filled and, therefore, for two interacting impurities 
the virtual hopping is allowed only in the AFM state making it the ground state \cite{zunger}.
For two Pb interstitials placed on distance 
4.1 \AA\ we found that $E_{\operatorname{M}}$=-0.96 eV while for distance 
12.5 \AA\ it drops down to -0.0056 eV. For the C interstitial the 
interactions are lower as $E_{\operatorname{M}}$=-0.38 eV
for distance 4.1 \AA\ and $E_{\operatorname{M}}$=-0.0023 eV for 12.5 \AA.
An application of HF for the localized electrons suppresses $E_{\operatorname{M}}$ almost twice.
The AFM interaction between impurities can be switched to FM 
for the impurity of $p^1$ occupation of outer shell (for example, In or Ga work for FM state) or
for $p^2$ occupation by choosing the charged state (1+).

In terms of the formation energies of the interstitials, it drastically drops down 
as the atomic radius decreases. Thus, if the formation energy of the Pb interstitial is 
1.23 eV for the Pb-rich/vacuum conditions, it 
is reduced almost to zero for Ge, Si impurities and already turns to be negative for the C and O impurity 
(the formation energy of the O interstital is -0.26 eV)
thus implying the higher solubility limit with prospects
of spontaneous defect appearance. 
Therefore, it is found that suppression in the magnetic coupling between impurities 
to occur with reduction of its atomic radius is compensated by simultaneous 
shift of the impurity solubility limit to the higher defect concentration.
For defects appearing on surface, 
the defect concentration can reach a number of sites available for doping ($\sim$ $10^{22}$ $cm^{-3}$)
as the formation energy is reduced further down 
(for example for the Pb interstitial it drops down by $\sim$ 1.0 eV such as the solubility limit is
$10^{20}$ $cm^{-3}$).

In summary, we propose to induce the $p$ local orbital magnetism
by doping of $\alpha$-PbO semiconductor with the non-magnetic impurities to appear in
layered structure of $\alpha$-PbO as the interstitial defects.
To create conditions for partially filled $p$ shell to act as $3d$ or $4f$ shells of the magnetic ions,
i.e. obeying the Hund's rule for the on-site ordering of the unpaired electrons,
an impurity of specific $s^2p^{1\leq x\leq 3}$ outer shell is required.
The combination of the outer shell with the
PbO crystal structure is unique because it allows doping to be
almost 'non-invasive' to the host lattice and its electronic properties.
In particular, bonding between impurity and the host 
involves only their $s^2$ outer shell electrons thereby 
preserving the original electronic configuration of the $p$ shells. 
In this case, the partially filled $p^{1\leq x\leq 3}$-shell of impurity
generates the localized spins on-site of defect.  
The main advantage of the defect-induced magnetism
over the diluted magnetic semiconductors is duality in its state localization. 
Thus, the $p$-localized state formed on site of the Pb interstitial has the localized nature to form the stable local magnetic moment,
but impurity-host hybridization results in the extended defect tails. 
The manifesting long-range order interactions between defects in combination with their high solubility 
create conditions for magnetic percolation to occur.\\

\section{Acknowledgement}
We would like to thank Prof. P. Fulde, 
Prof. T. Chakraborty and Dr. L. Hozoi 
for their guidance and thoughtful insights in our work. 
This work was made possible by the computational facilities of Dr. O. Rubel and 
the Shared Hierarchical Academic Research Computing Network (SHARCNET:www.sharcnet.ca)
and Compute/Calcul Canada.
Financial support of Ontario Ministry of Research and Innovation through a 
Research Excellence Program Ontario network for advanced medical imaging detectors is highly acknowledged.


\begin{thebibliography}{99}
\bibitem{exp_my}
Preliminary results: EPR measurements performed on PbO samples at room temperature have indicated an appearance 
of several paramagnetic centers.
\end{thebibliography}
\end{document}